\documentclass[aps,pre,preprint,showpacs,unsortedaddress,
superscriptaddress]{revtex4}
\usepackage{epsfig}
\usepackage{latexsym}
\begin{document}
\title{Hybrid lattice Boltzmann model for binary
fluid mixtures}
\author{A. Tiribocchi}
\email[]{adriano.tiribocchi@ba.infn.it} \affiliation{Dipartimento di
Fisica, Universit\`{a} di Bari,
 {\it and} INFN, Sezione di Bari,
Via Amendola 173, 70126 Bari, Italy}
\author{N. Stella}
\email[]{nicolastella1@gmail.com} \affiliation{Dipartimento di
Fisica, Universit\`{a} di Bari,
Via Amendola 173, 70126 Bari, Italy}
\author{G. Gonnella}
\email[]{gonnella@ba.infn.it} \affiliation{Dipartimento di
Fisica, Universit\`{a} di Bari,
 {\it and} INFN, Sezione di Bari,
Via Amendola 173, 70126 Bari, Italy}
\author{A. Lamura}
\email[]{a.lamura@ba.iac.cnr.it}
\affiliation{
Istituto Applicazioni Calcolo, CNR,
Via Amendola 122/D, 70126 Bari, Italy}
\date{\today}
\begin{abstract}
A hybrid lattice Boltzmann method (LBM) for binary mixtures based on
 the free-energy approach is proposed. Non-ideal terms of the
pressure tensor are included as a body force in the LBM  kinetic
equations, used to simulate the continuity and Navier-Stokes equations.
The convection-diffusion equation is studied by finite difference
methods.
Differential operators are discretized in order to reduce the
magnitude of spurious velocities. The algorithm has been shown to be
stable and reproducing the correct equilibrium behavior in simple
test configurations and to be Galilean invariant. 
Spurious velocities can be reduced of about an
order of magnitude with respect to standard discretization
procedure.
\end{abstract}
\maketitle

\section{Introduction}

In recent years lattice Boltzmann methods (LBM) \cite{LBM} have been widely
used to study multiphase fluids \cite{DUN}. Examples of applications are the
analysis of growth regimes in phase separation of binary mixtures \cite{CATES}
or
the study of backflow effects in liquid crystal behavior \cite{YEOLIQ}.
The  LBM approach is well suited for
dealing with complex geometries or for parallel implementations \cite{LBM}.
Moreover, in the free-energy approach \cite{YEOREV}, 
the mesoscale properties of
the fluid (interface structures, coupling with local order
parameters, etc.) can be straightforwardly inserted in the LBM
numerical scheme and taken under control. Due to the relevance of
the method, it is worth to further develop LBM algorithms in order
to improve numerical stability and accuracy, also by optimizing the
use of computer resources.

LBM dynamics is defined in terms of kinetic equations for a set of
populations $f_i$ representing, at each lattice site and time, the
density of particles moving in one of the allowed directions of a
given lattice. The sum over the directions $i$  of $f_i$ is  the
local density of the fluid while the first momentum is related to
the local fluid momentum. 
In one approach a forcing term is included
in the kinetic equations representing the interactions between the
components of the mixture \cite{SHAN}.  
Differently, the free-energy method was
originally developed by fixing  the second moment of the populations
in terms of the pressure tensor of the fluid mixture \cite{OSY}.
It has been applied
to complex fluids in Refs.~\cite{GONN,THREE,YEONEM}.

In this paper we consider an approach similar to the one of 
Ref.~\cite{WAGNER2}
where a free-energy dependent term is added as a body force in the
kinetic equations. This approach traces back to the work of
Guo {\it et al.} \cite{GUO} where a comparison with different
methods to introduce the force is reported.
With respect to the algorithm of Ref.~\cite{OSY}, 
this allows a better control of the continuum limit
still keeping all the advantages of the free-energy method. 
In Ref.~\cite{WAGNER2} a lattice Boltzmann equation is considered
for each component. 
Here we consider a ``hybrid''  algorithm where LBM is used to simulate
Navier-Stokes equations while finite-difference methods are
implemented to simulate the convection-diffusion equation. 
Such hybrid codes have been used for complex fluids \cite{xu}, 
liquid crystals \cite{maren} and thermal flows \cite{lall}.
This allows to  reduce in a relevant way  the amount of required memory
in systems with multi-component  order parameters or in simulations
of three-dimensional systems. 

A typical undesired effect due to discretization is 
the appearing of unphysical flow close to the interfaces. This flow,
often known as spurious velocities, 
can severely affect the quality of LBM simulations. In this work we
discretize the  differential operators by a procedure optimized for
reducing  the magnitude of spurious velocities, following the
so-called  ``stencil''  method applied in Ref.~\cite{POOL} to  a multiphase
one-component fluid. Here we will see that this method  
allows to
reduce spurious velocities of about an order of magnitude.

The paper is organized as follows. In the next section the LBM
algorithm proposed is described and details on the numerical
implementation are given. In Section 3 results of simulations
of test configurations are shown. We will see how spurious velocities
around curved interfaces can be reduced applying a more general stencil 
to discretize derivatives.
We will also discuss the convection of a drop under a constant force acting
for a finite time interval.
Then some conclusions will follow in Section 4.

\begin{center}
\section{The model}
\end{center}

The equilibrium properties of the fluid mixture
can be described by a free energy
\begin{equation}\label{freenergy}
{\cal {F}}= \int d{\bf r}\left[n T \ln n + \frac{a}{2}\varphi^{2}
+\frac{b}{4}\varphi^{4}+\frac{\kappa}{2}(\nabla\varphi)^{2}\right]
\end{equation}
where  $T$ is the temperature, 
$n$ is the total density of the mixture, and $\varphi$ is the
scalar order parameter representing the concentration difference
between the two components of the mixture. The term depending on $n$ gives rise
to the ideal gas pressure $p^i = n T$ which does not affect the phase
behavior. The terms in $\varphi$ in the free-energy density $f(n,\varphi, T)$
correspond to the typical
expression of Ginzburg-Landau free energy used in studies of phase
separation \cite{BRAY}.
The terms in the free energy can be distinguished in two parts: The
polynomial terms describe the bulk properties of the mixture
and the gradient term is related to the interfacial ones.

In the bulk terms the parameter $b$ is always positive to ensure
stability while the parameter $a$ can distinguish a disordered
($a>0$) and an ordered ($a<0$) mixture, in which the two components
coexist with equilibrium values 
$\pm \varphi_{eq}$ where 
$\displaystyle \varphi_{eq}=\sqrt{\frac{-a}{b}}$ \cite{ROWI}.
The equilibrium profile between the two coexisting bulk components is
\begin{equation} 
\varphi(x)=\varphi_{eq}\tanh (\frac{2x}{\xi})
\label{prof}
\end{equation} 
with
interface width
\begin{equation}\label{intwidth}
\xi=2\sqrt{\frac{2\kappa}{-a}}
\end{equation}
and surface tension
\begin{equation}\label{surftens}
\sigma=\frac{2}{3}\sqrt{\frac{2a^2\kappa}{b}}.
\end{equation}
The thermodynamic functions can be obtained from the free energy
(\ref{freenergy}) by differentiation. The chemical potential difference
between the two components is given by
\begin{equation}\label{chempot}
\mu=\frac{\delta{\cal F}}{\delta \varphi}=a\varphi +b\varphi^3 -
\kappa\nabla^2\varphi.
\end{equation}
The pressure $P_{\alpha\beta}$ is a tensor since interfaces in the fluid can
exert nonisotropic forces \cite{YAFLE}. The diagonal part $p_0$ can be
obtained from Eq.~(\ref{freenergy}) as
\begin{equation}\label{diagpress}
p_0=n\frac{\delta{\cal F}}{\delta n}+
\varphi\frac{\delta{\cal F}}{\delta\varphi}
-f(n,\varphi, T)= p^i +\frac{a}{2}\varphi^2+\frac{3b}{4}\varphi^4-
\kappa\varphi(\nabla^2\varphi)-\frac{\kappa}{2}(\nabla \varphi)^2.
\end{equation}
For a fluid with concentration gradients  $P_{\alpha\beta}$ 
has to verify the  general equilibrium condition 
$\partial_{\alpha}P_{\alpha\beta}=0$ \cite{EVANS}. A suitable
choice for the pressure tensor is
\begin{equation}\label{presstens}
P_{\alpha\beta}=p_0\delta_{\alpha\beta}+\kappa\partial_{\alpha}\varphi
\partial_{\beta}\varphi.
\end{equation}

The hydrodynamic equations of fluids follow from the conservation
laws for mass and momentum. For binary mixtures at constant
temperature the evolution of density, velocity and concentration
fields is described by the continuity, the Navier-Stokes and the
convection-diffusion equations \cite{GROOTMAZUR}, respectively,
\begin{eqnarray}
&&\partial_tn+\partial_{\alpha}(n u_{\alpha})=0 ,\label{conteqn}\\
\partial_t(n u_{\beta})&+&\partial_{\alpha}(nu_{\alpha}u_{\beta})=
-\partial_{\alpha} P_{\alpha\beta}
+\partial_{\alpha}\{\eta(\partial_{\alpha}u_{\beta}
+\partial_{\beta}u_{\alpha}-\frac{2\delta_{\alpha\beta}}{d}\partial_{\gamma}
u_{\gamma})+\zeta\delta_{\alpha\beta}\partial_{\gamma}u_{\gamma}\}=\nonumber \\
&=& -\partial_{\beta} (p^i) - \varphi \partial_{\beta} \mu
+\partial_{\alpha}\{\eta(\partial_{\alpha}u_{\beta}
+\partial_{\beta}u_{\alpha}-\frac{2\delta_{\alpha\beta}}{d}\partial_{\gamma}
u_{\gamma})+\zeta\delta_{\alpha\beta}\partial_{\gamma}u_{\gamma}\} ,
\label{NavStokeqn}\\
&&\partial_t\varphi + \partial_{\alpha}(\varphi
u_{\alpha})=\Gamma\nabla^2\mu ,\label{convdiffeqn}
\end{eqnarray}
where $\eta$ and $\zeta$ are the shear and the bulk viscosities,
$\Gamma$ is
the mobility coefficient, and $d$ is the dimensionality of the system.

Equations (\ref{conteqn})-(\ref{convdiffeqn}) can be
solved numerically. We use a mixed approach that consists of a
finite difference scheme for solving Eq.~(\ref{convdiffeqn}) and of a 
LBM approach with forcing term for
Eqs.~(\ref{conteqn}) and (\ref{NavStokeqn}). 
This has the advantage that the amount of required memory can be 
decreased so that larger systems can be simulated. 
In our case of
study, for a two-dimensional model on a square lattice with nine
velocities ($D2Q9$), this method
allows to reduce the required memory of $\sim 27\%$.
Actually, the convection-diffusion equation could have also been 
solved on a $D2Q5$
lattice \cite{rasin} and in this case the reduction in memory would have been
of $\sim 17\%$.
Moreover, the
spurious terms in the continuum equations
found in previous formulations based on a free energy \cite{OSY} can be
avoided.

\subsection{The Lattice Boltzmann scheme with forcing term}

To solve Eqs.~(\ref{conteqn}) and (\ref{NavStokeqn})
we use a Lattice
Boltzmann scheme on a lattice of size $L_x \times L_y$
in which each site
is connected to nearest and next-to-nearest neighbors. This is one of the
simplest geometries which reproduce correctly
the
Navier-Stokes equations in continuum limit and is shown in Fig.~1. 
Horizontal and vertical links
have length $\Delta x$ and diagonal links $\sqrt{2}\Delta x$. 
On each site
{\bf r} nine lattice velocity vectors ${\bf e}_i$ are defined. They
have modulus
$\displaystyle |{\bf e}_i|=\frac{\Delta x}{\Delta t_{LB}}\equiv c$, 
being $\Delta t_{LB}$
the time step, for $i=1, 2, 3, 4$  and modulus $|{\bf e}_i|=\sqrt{2}c$ for
$i=5, 6, 7, 8$.
Moreover, the zero velocity vector ${\bf e}_0={\bf 0}$ is defined.
A set of distribution function $\{f_i({\bf r},t)\}$ is defined on each
lattice site ${\bf r}$ at each time $t$.

In the LB scheme for simple fluids \cite{LBM} 
the distribution functions evolve during the time step $\Delta t_{LB}$ 
according to
a single relaxation-time Boltzmann equation \cite{BATH}
\begin{equation}\label{evoleqn}
f_{i}({\bf r}+{\bf e}_{i}\Delta t_{LB}, t+\Delta t_{LB})-f_{i}({\bf r}, t)=   
-\frac{\Delta t_{LB}}{\tau}[f_{i}({\bf r}, t)-f_{i}^{eq}({\bf r}, t)] ,
\end{equation}
where $\tau$ is a relaxation parameter and $f_i^{eq}({\bf r},t)$
are the local equilibrium distribution functions. 
The total density $n$ and the fluid momentum $n {\bf u}$ 
are defined by the following relations
\begin{equation}
n=\sum_if_i , \hspace{1.3cm} n {\bf u}=\sum_i f_i {\bf e}_i ,
\end{equation}
where ${\bf u}$ is the fluid velocity. The form of $f_i^{eq}$ must be chosen
so that the mass and momentum are locally conserved in each collision step,
therefore the following relations must be satisfied
\begin{eqnarray}
\sum_i(f_i^{eq}-f_i)=&0&\Rightarrow \sum_if_i^{eq}=n\label{consmass} ,\\
\sum_i(f_i^{eq}-f_i) {\bf e}_i=&{\bf 0}&\Rightarrow\label{consimp} 
\sum_if_i^{eq}{\bf e}_i=n{\bf u} .
\end{eqnarray}
Moreover,  the $f_i^{eq}$'s need to have some symmetries so that 
the Navier-Stokes equations are reproduced in the continuum limit.
A convenient choice 
for the local equilibrium distribution functions of an ideal fluid
in the case of a $D2Q9$ model is given by a second order
expansion in the fluid velocity ${\bf u}$ of the Mawwell-Boltzmann
distribution \cite{QIAN}
\begin{equation}\label{espdistrfunc}
f_i^{eq}({\bf r},t)=\omega_in\left[1+\frac{{\bf e}_i \cdot{\bf u}}{c^2_s}+
\frac{{\bf uu:}({\bf e}_i{\bf e}_i-c^2_s{\bf I})}{2c^4_s}\right],
\end{equation}
where $c_s=c/\sqrt{3}$ is the sound speed in this model, 
{\bf I} is the unitary 
matrix and a suitable choice for the coefficients $\omega_i$ is $\omega_0=4/9$,
$\omega_i=1/9$ for $i=1-4$, $\omega_i=1/36$ for $i=5-8$.
This form is such that
\begin{equation}\label{secmoment}
\sum_if_i^{eq} e_{i\alpha} e_{i\beta}=n c_s^2 \delta_{\alpha \beta}
+ n u_{\alpha} u_{\beta} .
\end{equation}

In order to simulate Eq.~(\ref{NavStokeqn}) where a nonideal pressure tensor 
$P_{\alpha \beta}$ appears, we adopt a LB model with a forcing term 
following a derivation similar to that of Ref.~\cite{GUO}.
In the case of Ref.~\cite{GUO} the model was used to study 
forced simple fluids while we address the case of a binary mixture with
interaction and interface contributions.
The evolution equation of the distribution
functions becomes
\begin{equation}\label{forceevoleqn}
f_{i}({\bf r}+{\bf e}_{i}\Delta t_{LB}, t+\Delta t_{LB})-f_{i}({\bf r}, t)=   
-\frac{\Delta t_{LB}}{\tau}[f_{i}({\bf r}, t)-f_{i}^{eq}({\bf r}, t)]
+\Delta t_{LB} F_i,
\end{equation} 
where $F_i$ is the forcing term to be properly determined. 
The equilibrium distribution functions 
(\ref{espdistrfunc}) are not changed except for the formal substitution 
${\bf u}\rightarrow {\bf u}^*$, where ${\bf u}^*$ is given by 
\begin{equation}
n{\bf u}^*=\sum_i f_i {\bf e}_i +\frac{1}{2}{\bf F}\Delta t_{LB} ,
\label{newvel}
\end{equation} 
${\bf F}$ being the force density acting on the fluid and ${\bf u}^*$ the
physical velocity. The expression of ${\bf F}$ for our case will be
given later.
The forcing term $F_i$ can be expressed as a power series at the second order
in the lattice velocity \cite{LADD}
\begin{equation}\label{forceterm}
F_i=\omega_i\left[A+\frac{{\bf B}\cdot{\bf e}_i}{c^2_s}+
\frac{{\bf C:}({\bf e}_i{\bf e}_i-c^2_s{\bf I})}{2c^4_s}\right] ,
\end{equation}
where $A$, ${\bf B}$, and ${\bf C}$ are functions of ${\bf F}$. The moments 
of the force verify
the following relations
\begin{equation}\label{momentsforceterm}
\sum_iF_i=A, \hspace{1cm} \sum_i F_i {\bf e}_i = {\bf B}, \hspace{1cm}
\sum_i F_i {\bf e}_i{\bf e}_i=c^2_sA{\bf I}+\frac{1}{2}[{\bf C}+{\bf C}^T] ,
\end{equation}
and have to be consistent with the hydrodynamic equations.

The continuum limit is obtained by 
using a
Chapman-Enskog expansion in the Knudsen number $\epsilon$:
\begin{equation}
f_i=f_i^{(0)}+\epsilon f_i^{(1)}+\epsilon^2f_i^{(2)}+.... ,\label{funcesp}
\end{equation}
\begin{equation}
\partial_t=\epsilon\partial_{t_{1}}+\epsilon^2\partial_{t_{2}} ,
\label{dertempesp}
\end{equation}
\begin{equation}
\partial_{{\bf r}}=\epsilon\partial_{{\bf r}_{1}} ,
\label{derspaceesp}
\end{equation}
\begin{equation}
{\bf F}=\epsilon{\bf F}_1,\hspace{0.5cm} A=\epsilon A_1, \hspace{0.5cm}
{\bf B}=\epsilon{\bf B}_1,\hspace{0.5cm} {\bf C}=\epsilon{\bf C}_1.
\label{forceesp}
\end{equation}
We note that the force term is
of first order in $\epsilon$ \cite{BUICK}. 
The continuity and the Navier-Stokes 
equations are recovered in the following form 
\begin{equation}
\partial_t(n u_{\beta}^*)+\partial_{\alpha}(nu_{\alpha}^*u_{\beta}^*)=
-\partial_{\beta} (n c_s^2) + F_{\beta} 
+\partial_{\alpha}\{\eta(\partial_{\alpha}u_{\beta}^*
+\partial_{\beta}u_{\alpha}^*)\}
\label{nav2}
\end{equation}
in terms of the velocity  ${\bf u}^*$ 
when  the 
following expressions for the terms $A$, ${\bf B}$, ${\bf C}$:
\begin{equation}
A=0, \hspace{1cm} {\bf B}=\left(1-\frac{\Delta t_{LB}}{2\tau}\right){\bf F},
\hspace{1cm} {\bf C}=\left(1-\frac{\Delta t_{LB}}{2\tau}\right)
({\bf u}^*{\bf F}+{\bf F} {\bf u}^*) 
\label{coeff}
\end{equation}
are used. The continuum equations (\ref{conteqn}) and (\ref{nav2})
can be  also obtained by a Taylor expansion method.
We remark that no spurious terms are present in the continuum equations
except for a term of order $u^{*3}$ which is neglected in Eq.~(\ref{nav2}).
Such approximation is correct as far as $u^{*2} << c_s^2$  when
the expansion (\ref{espdistrfunc}) is valid \cite{LBM}.
In the present formulation the second moment of the equilibrium
distribution function (\ref{secmoment}) does not need to be
modified to include the effects of the pressure tensor as in previous
models based on a free energy \cite{OSY}.
It is straightforward to show that the momentum defined
in Eq.~(\ref{newvel}) corresponds to an average between the pre- and 
post-collisional values of the velocity ${\bf u}$
which is the correct way to calculate it when a forcing
term is introduced \cite{SHAN,BUICK}. It is this value that
appears in the continuum equations and is measured in simulations.
As in the case of standard LBM \cite{LBM}, the present model is characterized
by the fact that $\displaystyle \zeta=\frac{2}{d} \eta$ with shear viscosity
\begin{equation}
\eta = n c_s^2 \Delta t_{LB}\left(\frac{\tau}{\Delta t_{LB}} 
- \frac{1}{2}\right). 
\end{equation}
In order to recover Eq.~(\ref{NavStokeqn}) we have to require that
\begin{equation}
{\bf F} = {\bf \nabla} (n c_s^2 - p^i) 
-\varphi {\bf \nabla} \mu = -\varphi {\bf \nabla} \mu .
\label{forza}
\end{equation}
The last equality comes from the fact 
the term $n c_s^2$ corresponds in LBM to the ideal gas pressure $p^i$
\cite{LBM}.
Finally, the forcing term in Eq.~(\ref{forceevoleqn}) has the form 
\begin{equation}\label{latticeforceterm}
F_i=\left(1-\frac{\Delta t_{LB}}{2\tau}\right)\omega_i\left[
\frac{{\bf e}_i-{\bf u}^*}{c^2_s}+\frac{{\bf e}_i\cdot{\bf u}^*}{c^4_s}
{\bf e}_i\right]\cdot {\bf F} 
\end{equation}
with ${\bf u}^*$ given by Eq.~(\ref{newvel}).

\subsection{Numerical calculation of the forcing term}

The derivatives of the order parameter in the forcing term (\ref{forza})
are calculated using
a finite difference scheme. In particular, we have adopted a stencil 
representation of finite difference operators in the more general way
to ensure higher isotropy \cite{POOL}, which is known to reduce
spurious velocities \cite{SHAN2,SUCCI2}.
The schemes for the $x$ derivative and the Laplacian operators
are, respectively,
\begin{equation}
\partial_{Dx}  = 
\frac{1}{\Delta x}
\left[
\begin{array}{ccc}
-M & 0 & M \\
-N & 0 & N \\
-M & 0 & M \\
\end{array}
\right] 
\label{xstencil}
\end{equation} 
\begin{equation}
\nabla_D^2   = 
\frac{1}{{\Delta x}^2}
\left[
\begin{array}{ccc}
R & Q & R \\
Q & -4\left(Q+R\right) & Q \\
R & Q & R \\
\end{array}
\right]
\label{laplstencil}
\end{equation}
with $2N+4M=1$ and $Q+2R=1$ to guarantee consistency between the continuous and
discrete derivatives \cite{POOL}. 
The subscript $D$ in the symbols of derivatives 
denotes 
the discrete operator. In these schemes the central entry is referred to
the lattice point where the derivative is computed, and the other entries are
referred to the eight neighbor lattice sites. The discrete derivatives of the
order parameter $\varphi$ are computed by summing the values in the
site and in the eight neighbors with the weights in the matrices 
(\ref{xstencil})-(\ref{laplstencil}).
The $y$ derivative is computed by transposing the matrix (\ref{xstencil}). 
The choice of the free
parameters $N$ and $Q$ is made in such a way 
that the spurious velocities are minimized (see next Section). 
We will refer to this case as the optimal choice (OC).
The values $N=1/2$, $M=0$, 
$Q=1$, and $R=0$ correspond to the standard central difference scheme denoted
as SC.
We will compare SC and OC in the following.

\subsection{The scheme for the convection-diffusion equation}

The convection-diffusion equation (\ref{convdiffeqn})
is solved by using 
a finite difference scheme.
The function $\varphi({\bf r},t)$ is defined on the nodes of the same lattice
used for the LB scheme. The time is discretized in 
time steps $\Delta t_{FD}$ with time values
$t^n=n \Delta t_{FD}$, $n=1,2,3,...$.
The relationship connecting the two time steps is
$\Delta t_{LB}= m \Delta t_{FD}$, being $m$ an integer.
We denote any discretized function at time $t^n$ 
on a node $(x_i,y_j)$ ($i=1,2,...,L_x; j=1,2,...,L_y$) 
of the lattice
by $g(x_i,y_j, t^n)= g_{ij}^n$. At each time step we update 
$\varphi^n \rightarrow \varphi^{n+1}$ using Eq.~(\ref{convdiffeqn})
in two successive partial steps \cite{FIELDING}.
This allows to have a better numerical stability.
In the first step we implement the convective term using an explicit
Euler algorithm \cite{STRIK}
\begin{equation}
\varphi^{n+1/2}=\varphi^n 
- \Delta t_{FD} (\varphi^n\partial_{\alpha}u_{\alpha}^{*n}+
u_{\alpha}^{*n}\partial_{\alpha}\varphi^n)
\label{adv}
\end{equation}
where the velocity ${\bf u}^*$ comes from the solution of the LB equation.
Note that the term $\partial_{\alpha}u_{\alpha}^{*n}$ has not been neglected
since the fluid is not exactly incompressible. Indeed,
the Navier-Stokes equation (\ref{nav2}) coming from the LBM contains
some compressibility terms which can be anyway kept very small requiring that
$u^{*2} << c_s^2$ \cite{LBM}. 
The 
derivatives in (\ref{adv}) are discretized as follows:
\begin{eqnarray}
\label{centrdiff}
\partial_{Dx} u^*_x|^n_{ij}&=&\frac{u^{*n}_{x,(i+1)j}
-u^{*n}_{x,(i-1)j}}{2\Delta x}  \\
\label{backdiff}
\partial_{Dx}\varphi|^n_{ij}&=&\frac{\varphi^n_{ij}-
\varphi^n_{(i-1)j}}{\Delta x}  \hspace{1 cm} \mbox{\rm if} 
\hspace{1 cm} u^{*n}_{x,ij} > 0
\\
\label{fordiff}
\partial_{Dx}\varphi|^n_{ij}&=&\frac{\varphi^n_{(i+1)j}-
\varphi^n_{ij}}{\Delta x}  \hspace{1 cm} \mbox{\rm if} 
\hspace{1 cm} u^{*n}_{x,ij} < 0
\end{eqnarray}
and analogously for the $y$ components.

The diffusive part of Eq.~(\ref{convdiffeqn}) is implemented in
the second update step using an explicit
Euler algorithm as
\begin{equation}
\varphi^{n+1}=\varphi^{n+1/2} 
+ \Delta t_{FD} \Gamma \left [a \nabla^2 \varphi^{n+1/2} + b \nabla^2 f^n 
- \kappa \nabla^2 (\nabla^2 \varphi^{n+1/2}) \right ]
\label{diff}
\end{equation}
where $f^n=(\varphi^n)^3$ and 
the operator $\nabla^2$ is discretized using the form given in 
Eq.~(\ref{laplstencil}) with the standard choice $Q=1$ and $R=0$.
Other choices using a more general stencil 
for discretizing $\nabla^2$ are possible though
we checked that they did not provide any relevant difference.

\section{Results and discussion}

We considered several test cases in order to validate our model.
We used the values $\Delta x= \Delta t_{LB}=\Delta t_{FD} = 1$. 
In the free energy we adopted the parameters
$-a = b = 10^{-3}$, $\kappa= - 3 a$ corresponding to an equilibrium
interface of width $\xi \simeq 5 \Delta x$. The mobility $\Gamma$ was set to 
5 and the relaxation time $\tau/\Delta t_{LB}$ was $1$ 
unless differently stated.

We first examined the relaxation to equilibrium of a planar sharp interface
on a lattice of size $L_x=L_y=64$ 
varying $\tau$ in the SC case. In all the cases
the system correctly relaxes to the expected profile (\ref{prof}).
One example is reported in
Fig.~2. 
In the case of a planar interface the fluid velocities $u^{*}$ decay to 
negligible values as it should be at equilibrium when $\Delta \mu = 0$ and
$\partial_{\alpha}P_{\alpha\beta}=0$.

We then studied a circular drop as a test for a case
with interfaces not aligned with the lattice links. 
A drop with sharp interface 
of diameter $64 \Delta x$ was placed at the center
of a lattice of size $L_x=L_y=128$ and let equilibrate in the SC case. 
Interfaces relax to the 
expected profile without deforming the drop but spurious velocities appear
as it can be seen in the upper panel
of Fig.~3 in the case with $\tau/\Delta t_{LB}=5$. 
We then used the OC scheme
to verify whether spurious velocities could be reduced by using a more 
isotropic structure for the discrete spatial derivatives in the forcing
term (\ref{forza}). We scanned several values of $N$ and $Q$ in order
to reduce the maximum value of the velocity $|u^{*}_{max}|$ on the whole
lattice. The optimal values are summarized in the Table I.
It is interesting to note that there is a couple of values
$N=0.3$ and $Q=2.5$ which occurs more frequently. We verified that this choice
is also effective in reducing spurious velocities even for the other values
of $\tau$. For this choice of $N$ and $Q$ the maximum velocities differ
only by a small percentage from the tabled values.
\begin{table}[h]
{\begin{tabular}{@{}cccc@{}} \toprule
$\tau/\Delta t_{LB}$ \hphantom{10} & $N$ \hphantom{10} & $Q$ \hphantom{10} & 
$|u^{*}_{max}|/c_s$ \hphantom{10} \\ \colrule
0.6 \hphantom{10} & 0.3 \hphantom{10} & 3 \hphantom{10} & 0.0001753 
\hphantom{10}  \\
0.8 \hphantom{10} & 0.3 \hphantom{10} & 2.5 \hphantom{10} & 0.0000603 
\hphantom{10}  \\
1 \hphantom{10} & 0.3 \hphantom{10} & 2.5 \hphantom{10} & 0.0000365 
\hphantom{10}  \\
1.2 \hphantom{10} & 0.3 \hphantom{10} & 2.5 \hphantom{10} & 0.0000267 
\hphantom{10}  \\
5 \hphantom{10}  & 0.3 \hphantom{10} & 2.5 \hphantom{10} & 0.0000088 
\hphantom{10}  \\
10 \hphantom{10} & 0.3 \hphantom{10} & 2 \hphantom{10} & 0.0000062 
\hphantom{10}  \\ \botrule
\end{tabular} \label{tbl:scaling}}
\caption{Optimal values of $N$ and $Q$ for different values of $\tau$ and the
corresponding values of the maximum spurious velocity $|u^{*}_{max}|$.}
\end{table}
Velocities can be greatly reduced with respect to the SC case
as it can be visually observed in the lower
panel of Fig.~3.

We also tried to get an analytical estimate of the optimal values of $N$ and
$Q$ in the following way. At
equilibrium it holds that $\partial_{\alpha} P_{\alpha\beta} = \varphi
\partial_{\beta} \mu = (a \varphi + 3 b
\varphi^3) \partial_{\beta} \varphi - k \varphi \partial_{\beta} (\nabla^2
\varphi) = 0$. 
This expression depends  on the first- and third-order derivatives.
By using the stencils (\ref{xstencil})-(\ref{laplstencil}) 
we get for the discretized operators the expressions
\begin{equation}
\partial_{Dx} = \partial_x
+ \frac{1}{6} (\Delta x)^2 \partial_x^3
+ \frac{1-2N}{2} (\Delta x)^2 \partial_x \partial_y^2 + ...
\end{equation}
and
\begin{equation} 
\nabla^2_D = \nabla^2
+ \frac{1}{12} (\Delta x)^2 (\partial_x^4 + \partial_y^4)
+ \frac{1-Q}{2} (\Delta x)^2 \partial_x^2 \partial_y^3 + ... ,
\end{equation}
so that
\begin{eqnarray}
\partial_{Dx} (\nabla^2_D) = \partial_x (\nabla^2)
+ \frac{1}{4} (\Delta x)^2 \partial_x^5
&+& \Big[\frac{1}{6} + \frac{1-2N}{2}
+ \frac{1-Q}{2}\Big] (\Delta x)^2 \partial_x^3 \partial_y^2\\ \nonumber
&+& \Big[\frac{1}{12} + \frac{1-2N}{2} \Big]
(\Delta x)^2 \partial_x \partial_y^4 + ...
\label{39}
\end{eqnarray}
By imposing that the error terms in the third-order derivative
depending on $N$ and $Q$ vanish, we get $N = 7/12 \simeq 0.6$ and $Q
= 7/6 \simeq 1.2$. However, this estimate does not correspond to the
optimal results of Table I. This is due to the fact that 
these optimal values were found by considering the full
dynamical problem with  the whole set of equations where we minimized the
spurious velocities. 
In the estimate after Eq.~(\ref{39}) the coupling with the velocity
field was not taken into account so that 
there is no {\it a priori} reason to expect the same
optimal values for $N$ and $Q$.

A comparison of the spurious velocities in the SC and OC cases 
is shown in Fig.~4. 
By using the optimal choice OC
the 
spurious velocities can be reduced by a factor approximately 10 with respect 
to the standard case SC over the whole range of $\tau$ values.
The stencil forms (\ref{xstencil})-(\ref{laplstencil}) 
were also applied to the model of Ref.~\cite{OSY} for 
nonideal fluids finding a comparable reduction in the magnitude of
spurious velocities with respect to the standard case \cite{POOL}. 

We then studied the motion of an equilibrated drop of
diameter $64 \Delta x$ in a lattice of
size $L_x=256, L_y=128$ under the effect
of an external constant force that acts up to the time $t/\Delta t_{LB}=500$
and is then switched off.
The additional force ${\bf G}=n(g_x,0) \Delta x^2/\Delta t_{LB}$
acts on the total density. $g_x$ is in the range
$[10^{-5} , 5 \times 10^{-5}]$ and the OC scheme
is used.  
The overall system is set in motion rightwards with increasing velocity until
the force ${\bf G}$ is on, then it moves with constant speed.
The choice of $g_x$ is such that the final velocity is
much smaller than the speed of sound $c_s$.
The aim was to check whether the system is Galilean invariant and the
drop is correctly convected by the flow. We monitored the shape
of the drop and measured its center of mass velocity ${\bf v}_{CM}$. 
This is defined as the
average velocity of the center of mass whose position is
\begin{equation}
{\bf r}_{CM}(t) = \frac{\sum_{ij} \varphi_{ij} {\bf r}_{ij}(t)}{\sum_{ij} 
\varphi_{ij}} 
\end{equation}
where the sum is over the lattice nodes ${\bf r}_{ij}$ inside the drop.
This velocity represents the convection velocity and is
compared with the fluid velocity 
${\bf v}_f (t) = {\bf u}^{*}({\bf r}_{CM}(t)) $ at the center of mass
given directly by the LBM.
In Fig.~5 the comparison between the velocities ${\bf v}_{CM}$ and
${\bf v}_f$ along the $x$-direction is shown
in the case with $g_x = 3 \times 10^{-5}$. It is evident that the two coincide
indicating that the drop is correctly advected by the fluid. Moreover, its
shape is not altered by motion as it can be seen
in Fig.~6 where some configurations of the system at different times 
are presented. Moreover, the drop is shown to make clear that it does 
not change in shape with time.
We measured the ratio of the horizontal and vertical diameters finding that
it stays almost constant with a deviation less than $3\%$ from the value 1. 
If the advection velocity is higher, the drop will be slightly deformed being
stretched along the $x$-direction. This effect becomes negligible when
increasing the surface tension (\ref{surftens}) via the parameter $\kappa$.

\section{Conclusions}

In this paper we have considered a lattice Boltzmann method for
binary mixtures with thermodynamics fixed by a free-energy
functional. We used a mixed method, with continuity and
Navier-Stokes equations simulated by LBM, and convection-diffusion
equation by finite difference schemes. Differently than in previous
free-energy LBM formulations \cite{OSY}, 
the interaction part in the
pressure tensor is not introduced by fixing  the second moment of
the LBM populations but by introducing a forcing term in the lattice
equation. 
This approach is
suggested by a microscopic picture and allows
to obtain a continuum limit without spurious terms. 
On the
other hand, the mixed or hybrid approach allows a reduction in the
required memory and this  can be relevant in performing large-scale
simulations.

In order to reduce spurious velocities, differential operators have
been  discretized by generalizing the usual lattice representations.
Free parameters appear and their optimal values have been fixed by
requiring that the maximum value of spurious velocities at
equilibrium is minimized.

We considered simple test situations, flat interfaces and single
drops showing that the correct equilibrium profiles are
reproduced. We found that  spurious velocities are reduced of about
an order of magnitude when a more general stencil is applied to the
derivatives  in the forcing term of the LBM equations. We did not
found any relevant difference by applying this procedure to the
differential operators appearing in the convection-diffusion
equation. We also checked that our method is stable in phase
separation studies, even if we have not reported the results of
these simulations in this work. Finally, we checked the effective
Galilean invariance of the system by advecting for some time
interval by a constant force a configuration with one drop and
then letting the system to evolve without forcing. For the cases
considered, we did not observe relevant drop deformations, the drop being
correctly advected by the surrounding fluid. 

In conclusion, we hope that this development of the free-energy LBM
can be useful in future simulations of binary mixtures and complex
fluids.

\newpage

\begin{figure}[ht]
\begin{center}
   \includegraphics[height=12cm,angle=0]{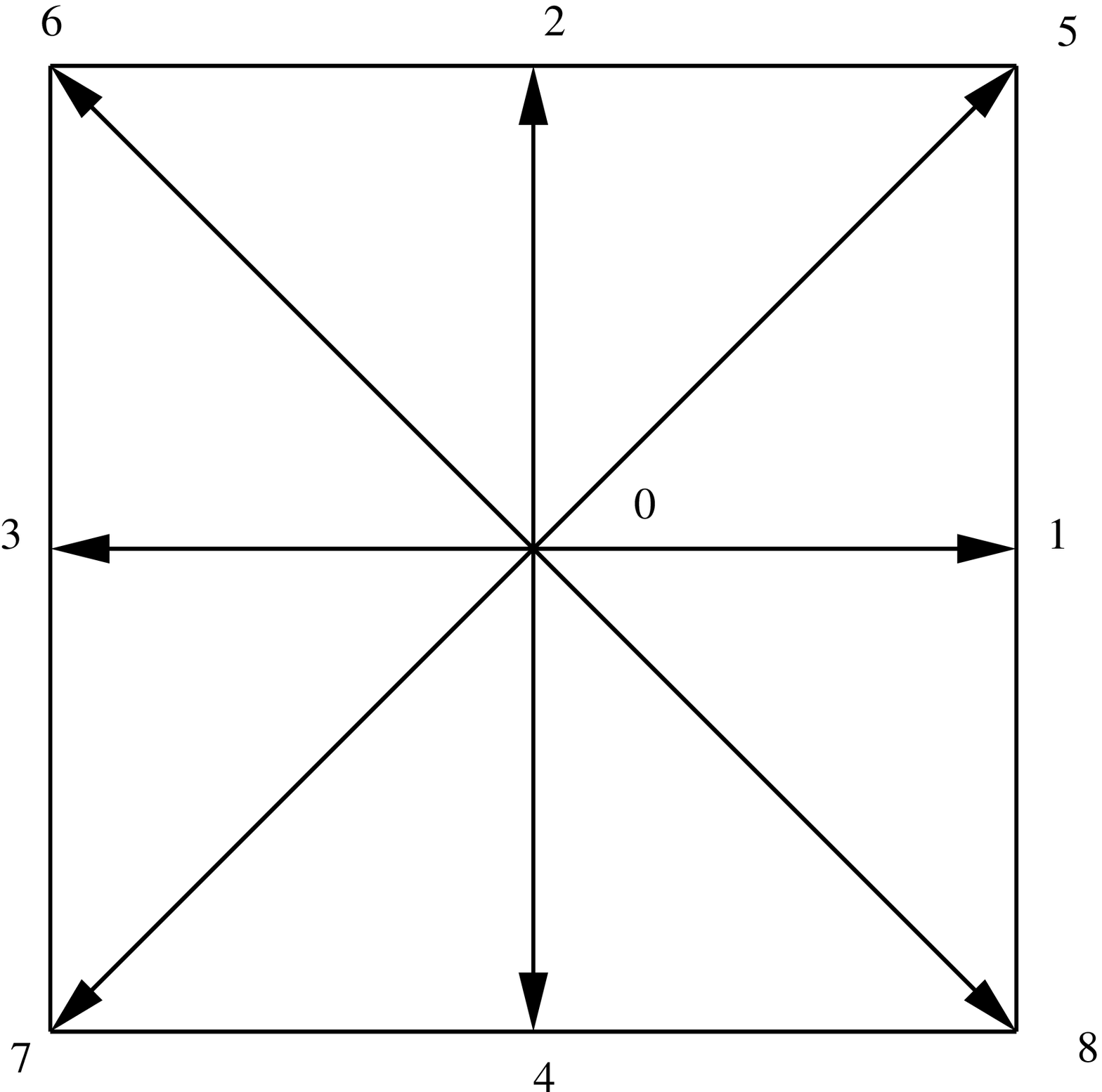}
\caption{Cell of the $D2Q9$ lattice used in the present study.
}
\end{center}
\label{figlat}
\end{figure}

\newpage

\begin{figure}[ht]
\begin{center}
   \includegraphics[height=12cm,angle=-90]{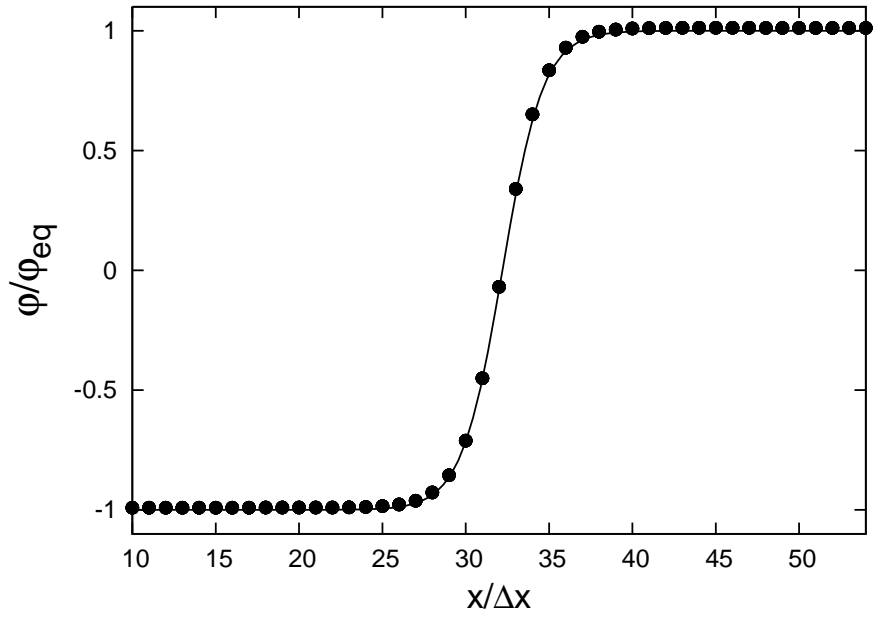}
\caption{Equilibrium profile of a planar interface on a lattice of size 
$L_x=L_y=64$
in the SC case. The continuous line is the
analytical result (\ref{prof}) and data points are the results of simulations.}
\end{center}
\label{figprof}
\end{figure}

\newpage

\begin{figure}[ht]
\begin{center}
  \includegraphics[height=12.7cm,angle=-90]{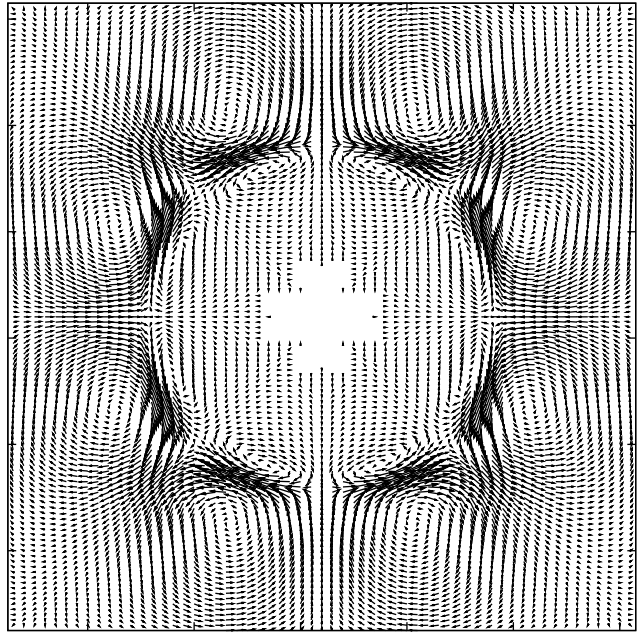}
\end{center}
\begin{center}
   \includegraphics[height=12.7cm,angle=-90]{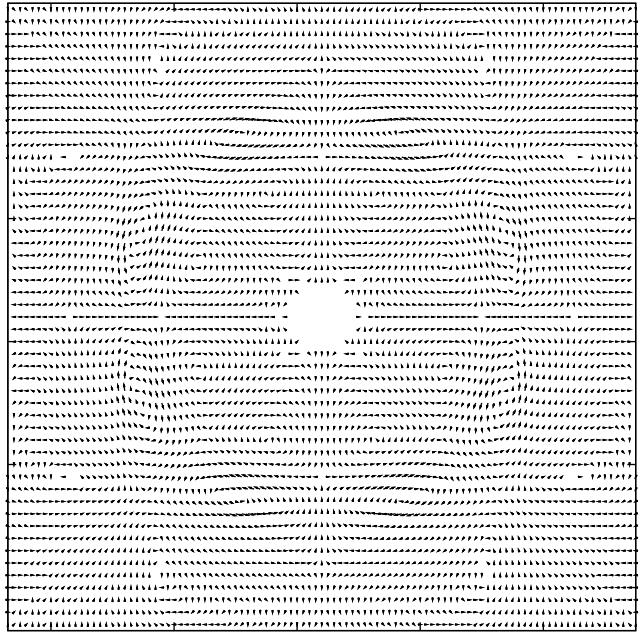}
\caption{Velocity patterns (the same scale is used in both the panels) 
at equilibrium when $\tau/\Delta t_{LB}=5$ in the
SC case (upper panel) and in the OC case (lower panel). Empty spaces are due 
to negligible values of velocity. In both the cases the system has
size $L_x=L_y=128$.}
\end{center}
\label{figspur}
\end{figure}

\newpage

\begin{figure}[ht]
\begin{center}
   \includegraphics[height=11cm,angle=0]{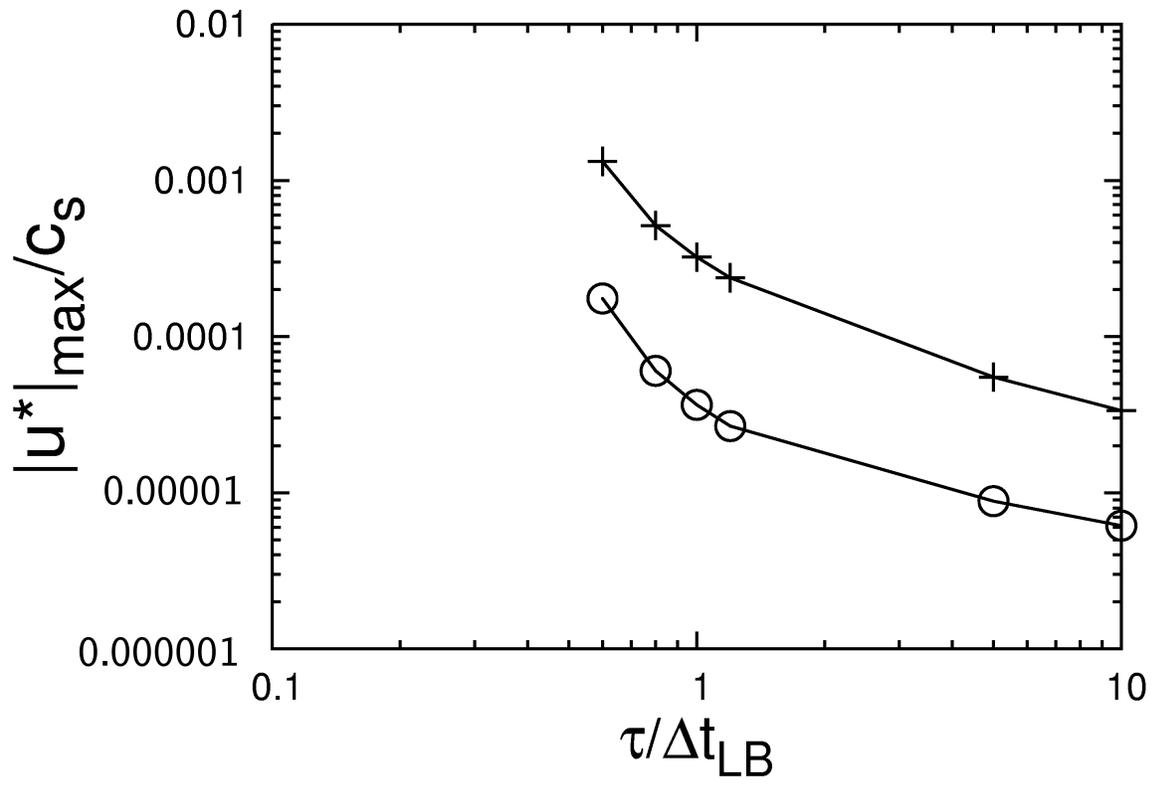}
\caption{Maximum spurious velocities as a function of $\tau$ 
in the SC case (+) and in the OC case ($\circ$).}
\end{center}
\label{figconfr}
\end{figure}

\newpage

\begin{figure}[ht]
\begin{center}
   \includegraphics[height=10cm]{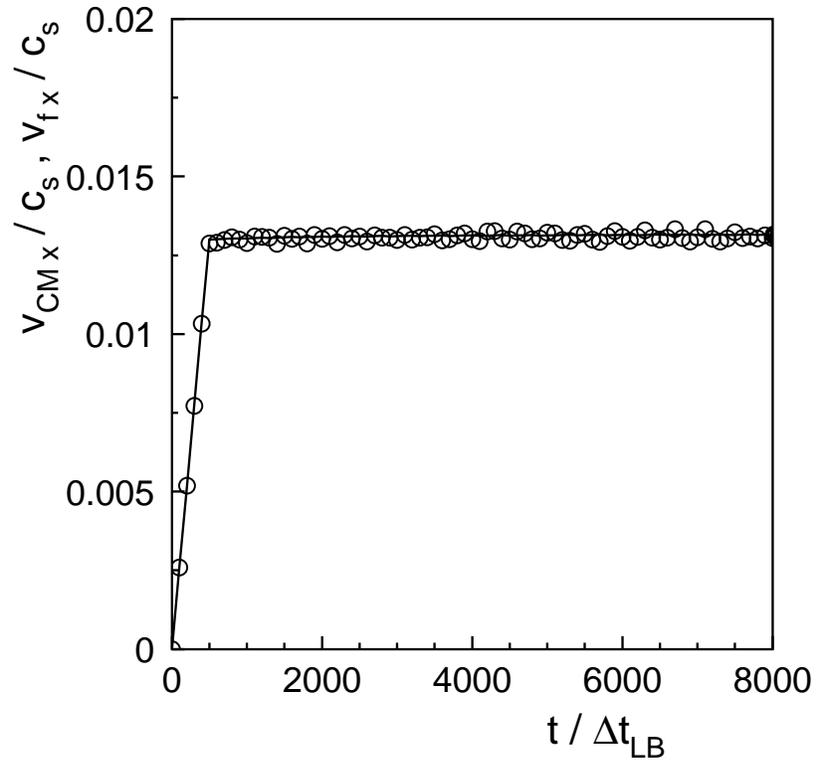}
\caption{Velocities of the center of mass of the drop $v_{CM x}$ ($\circ$)
and of the fluid  $v_{f x}$ ($-\!\!\!-\!\!\!-\!\!\!-$) at the center of mass 
along the $x$-direction as a function of time.
The external force $G_x$ acts until the time $t/\Delta t_{LB}=500$.}
\end{center}
\label{figadv}
\end{figure}

\newpage

\begin{figure}[ht]
\begin{center}
   \includegraphics[height=11cm]{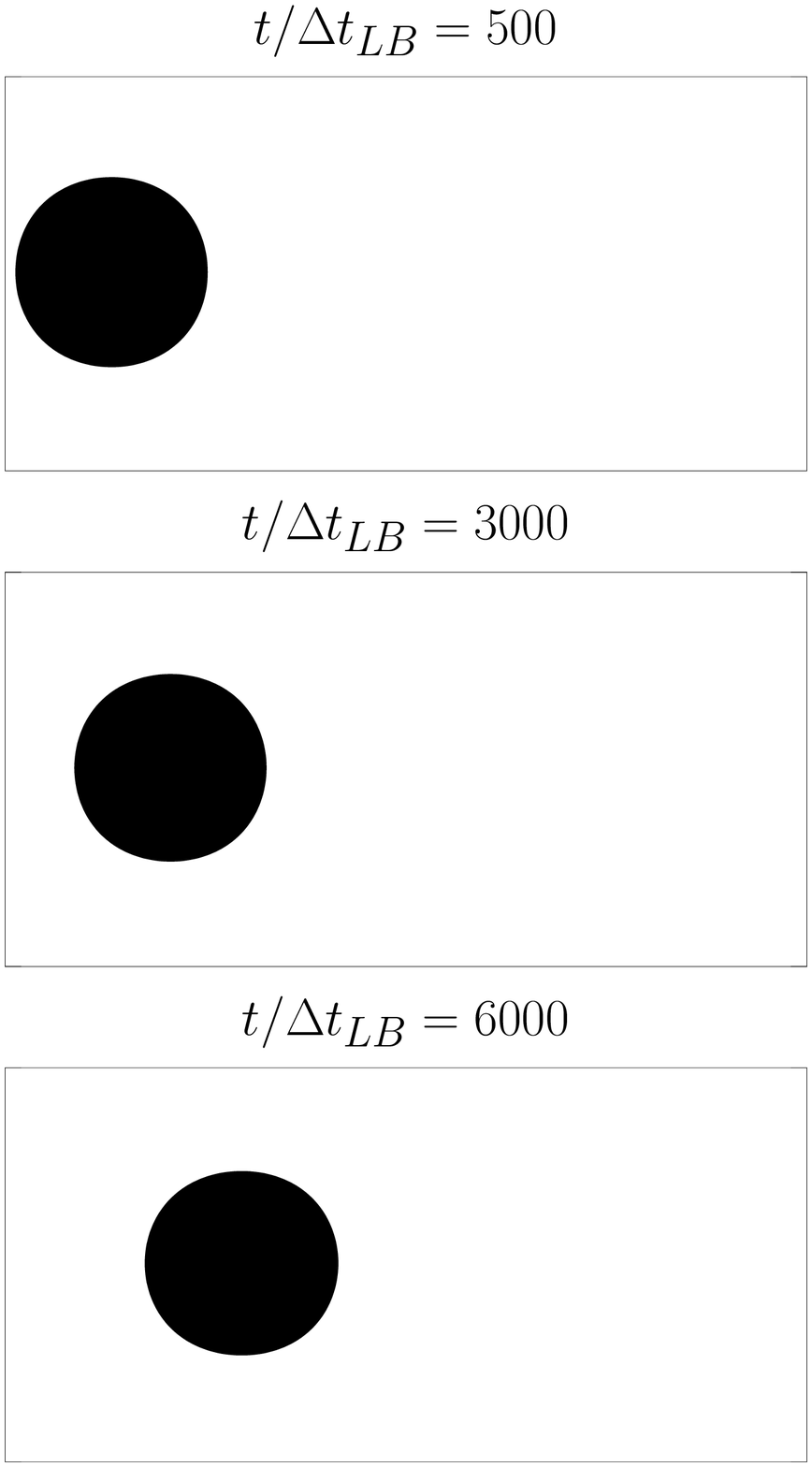}\\*
\includegraphics[height=9.5cm]{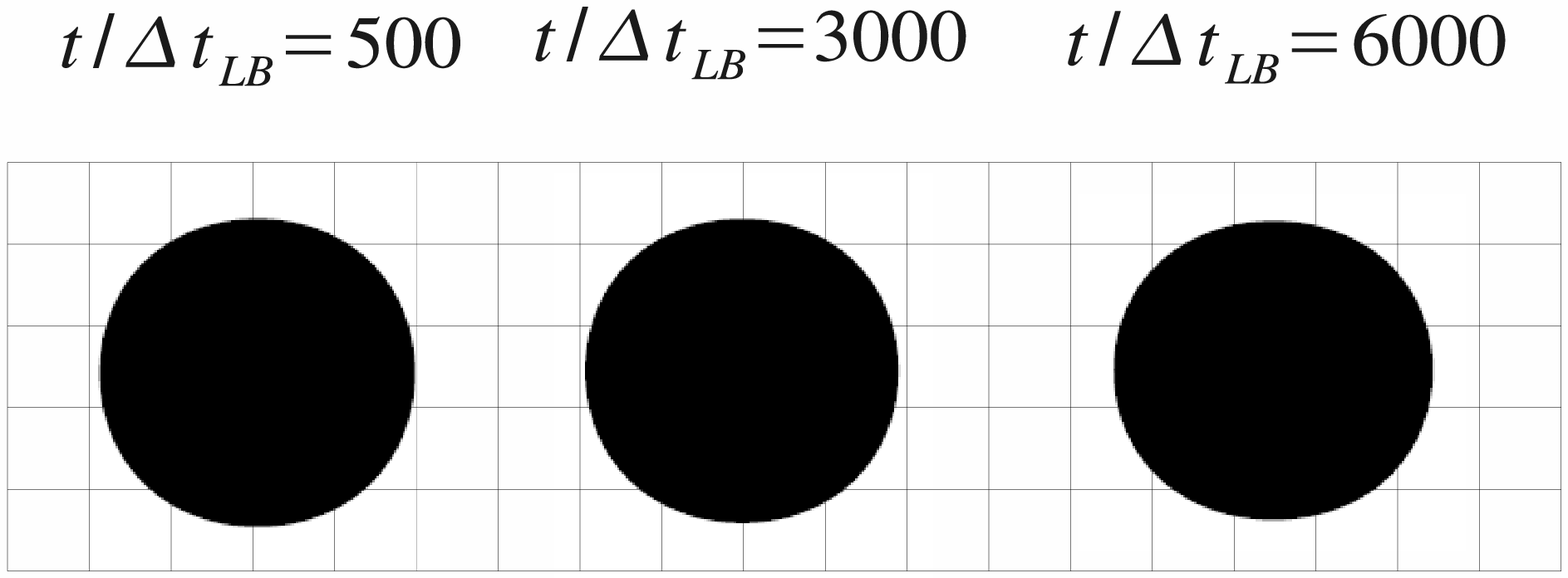}
\caption{Configurations of the advected drop at consecutive times.
The system has
size $L_x=256, L_y=128$.
In the lower panel the drop, extracted from the system, 
is shown on an underlying mesh to better appreciate its
shape.}
\end{center}
\label{figgoccia}
\end{figure}


\begin{thebibliography}{99}

\bibitem{LBM}
R. Benzi, S. Succi, and M. Vergassola, Phys. Rep. {\bf 222},
145 (1992);
S. Chen and G. D. Doolen, Annu. Rev. Fluid Mech. {\bf 30}, 329 (1998);
S. Succi, {\it The Lattice Boltzmann Equation for Fluid Dynamics and Beyond}
(Clarendon Press, Oxford, 2001).

\bibitem{DUN}
B. D\"{u}nweg and A. J. C. Ladd, Adv. Polym. Sci. {\bf 221}, 89 (2009).

\bibitem{CATES}
V. M. Kendon, J.-C. Desplat, P. Bladon, and M. E. Cates, 
Phys. Rev. Lett. {\bf 83}, 576 (1999).

\bibitem{YEOLIQ}
C. Denniston, E. Orlandini, and J. M. Yeomans,
Phys. Rev. E {\bf 63}, 056702 (2001).

\bibitem{YEOREV}
J. M. Yeomans, Annu. Rev. Comput. Phys. {\bf 7}, 61 (1999).

\bibitem{SHAN}
X. Shan and H. Chen, Phys. Rev. E {\bf 49}, 2941 (1994).

\bibitem{OSY}
E. Orlandini, M. R. Swift, and J. M. Yeomans, Europhys. Lett. {\bf 32}, 463
(1995); M. R. Swift, E. Orlandini, W. R. Osborn, and J. M. Yeomans, 
Phys. Rev. E {\bf 54}, 5041 (1996).

\bibitem{GONN}
G. Gonnella, E. Orlandini, and J. M. Yeomans, Phys. Rev. Lett. 
{\bf 78}, 1695 (1997).

\bibitem{THREE}
A. Lamura, G. Gonnella, and J. M. Yeomans, Europhys. Lett. {\bf 45}, 314
(1999).

\bibitem{YEONEM}
M. E. Cates, S. M. Fielding, D. Marenduzzo, E. Orlandini, 
and J. M. Yeomans,
Phys. Rev. Lett. {\bf 101}, 068102 (2008).

\bibitem{WAGNER2}
Q. Li and A. J. Wagner, Phys. Rev. E {\bf 76}, 036701 (2007).

\bibitem{GUO}
Z. Guo, C. Zheng, and B. Shi, Phys. Rev. E {\bf 65}, 046308 (2002).

\bibitem{xu}
A. G. Xu, G. Gonnella, and A. Lamura, Physica A {\bf 362}, 42
(2006).

\bibitem{maren}
D. Marenduzzo, E. Orlandini, M. E. Cates,
and J. M. Yeomans, Phys. Rev. E {\bf 76}, 031921 (2007).

\bibitem{lall}
P. Lallemand and L. S. Luo, Int. J. Mod. Phys. B {\bf 17}, 41 (2003);
F. Dubois and P. Lallemand, preprint, arXiv:0811.0599v2 [math.NA]. 

\bibitem{POOL}
C. M. Pooley and K. Furtado, Phys. Rev. E {\bf 77}, 046702 (2008).

\bibitem{BRAY}
A. J. Bray, Adv. Phys. {\bf 43}, 357 (1994).

\bibitem{ROWI}
J. S. Rowlinson and B. Widom, {\it Molecular Theory
of Capillarity} (Clarendon Press, Oxford, 1982).

\bibitem{YAFLE}
A. J. M. Yang, P. D. Fleming, and J. H. Gibbs, J. Chem. Phys. {\bf 64},
3732 (1976).

\bibitem{EVANS}
R. Evans, Adv. Phys. {\bf 28}, 143 (1979).

\bibitem{GROOTMAZUR}
S. R. De Groot and P. Mazur, {\it Non-equilibrium Thermodynamics}
(Dover Publications, New York, 1984).

\bibitem{rasin}
I. Rasin, S. Succi, and W. Miller, J. Comput. Phys. {\bf 206}, 453 (2005).

\bibitem{BATH}
P. Bathnagar, E. P. Gross, and M. K. Krook, Phys. Rev. {\bf 94}, 511 (1954).

\bibitem{QIAN}
Y. Qian, D. d'Humieres, and P. Lallemand, Europhys. Lett. {\bf 17}, 479 (1992).

\bibitem{LADD}
A. J. C. Ladd and R. Verberg, J. Stat. Phys. {\bf 104}, 1191 (2001).

\bibitem{BUICK}
J. M. Buick and C. A. Greated, Phys. Rev. E {\bf 61}, 5307 (2000).

\bibitem{SHAN2}
X. Shan, Phys. Rev. E {\bf 73}, 047701 (2006).

\bibitem{SUCCI2}
M. Sbragaglia, R. Benzi, L. Biferale, S. Succi, K. Sugiyama, and F. Toschi, 
Phys. Rev. E {\bf 75}, 026702 (2007).

\bibitem{FIELDING}
S. M. Fielding, Phys. Rev. E {\bf 77}, 021504 (2008).

\bibitem{STRIK}
J. C. Strikwerda, {\it Finite Difference Schemes and Partial Differential
Equations}
(Chapman \& Hall, New York, 1989).

\end{thebibliography}
\end{document}